# Why do the quantum observables form a Jordan operator algebra?

Gerd Niestegge
Zillertalstrasse 39, D-81373 Muenchen, Germany

*Abstract*. The Jordan algebra structure of the bounded real quantum observables was recognized already in the early days of quantum mechanics. While there are plausible reasons for most parts of this structure, the existence of the distributive non-associative multiplication operation is hard to justify from a physical or statistical point of view.

Considering the non-Boolean extension of classical probabilities, presented in a recent paper, it is shown in the present paper that such a multiplication operation can be derived from certain properties of the conditional probabilities and the observables, i.e. from postulates with a clear statistical interpretation.

The well-known close relation between Jordan operator algebras and C*-algebras then provides the connection to the quantum-mechanical Hilbert space formalism, thus resulting in a novel axiomatic approach to general quantum mechanics that includes the type II, III von Neumann algebras.

## 1. Introduction

A non-Boolean extension of classical probabilities was presented in [10]. The main purpose of [10] was to elaborate on the interpretation of the model and on some applications to quantum measurement. For that purpose, it was sufficient to consider finitely-additive probabilities.

In the present paper, countably-additive probabilities are needed to study observables which are defined in an abstract way as the analogue of the classical random variables. It is shown that, under certain conditions, a multiplication operation exists on the system of bounded real-valued observables which form a Jordan operator algebra then. The associativity of the multiplication operation is equivalent to the classical case. Since almost all Jordan operator algebras can be represented on a Hilbert space, the non-Boolean extension of the classical probabilities thus provides an axiomatic access to quantum mechanics.

Other axiomatic approaches to quantum mechanics start from different postulates including either an orthomodular partial ordering on the system of events (called propositions in the quantum logical approaches; e.g. [3,11,12,16]) or a distributive multiplication operation on the system of bounded real observables (e.g. [8,14,15]) or both (e.g. [13]). Postulating the existence of the distributive multiplication operation is hard to justify from a physical or statistical point of view (when using the so-called Segal product[13,14], the distributivity becomes a problem as pointed out in [15]), and the purely logical approaches are neither able to rule out some physically irrelevant cases [9] nor to cover the physically relevant type II,III von-Neumann algebras (which do not contain the minimal events needed for the geometrical methods these approaches use).

The approach of the present paper assumes a rather weak structure for the system of events (with a simple orthogonality relation instead of an orthomodular partial ordering); more important are certain statistically interpretable properties, postulated for the conditional probabilities and the observables, where the distributive multiplication operation can be derived from. This approach is closer to Kolmogorov's measure-theoretic access to classical probability theory than the other approaches. Moreover, the type II,III cases are included and the physically irrelevant cases discovered in [9] are excluded.

## 2. Non-Boolean probabilities

An *orthospace*[10] is a set $\mathcal{E}$ with distinguished elements 0 and $\mathbb{I}$, a relation $\perp$ and a partial binary operation + such that for $D,E,F \in \mathcal{E}$:

(OS1)   $E \perp F \Rightarrow F \perp E$
(OS2)   $E+F$ is defined for $E \perp F$, and $E+F=F+E$
(OS3)   $D \perp E,\ D \perp F,\ E \perp F \Rightarrow D \perp E+F,\ F \perp D+E$ and $D+(E+F)=(D+E)+F$
(OS4)   $0 \perp E$ and $E+0=E$ for all $E \in \mathcal{E}$
(OS5)   For every $E \in \mathcal{E}$ there exists a unique $E' \in \mathcal{E}$ such that $E \perp E'$ and $E+E'=\mathbb{I}$
(OS6)   $E \perp F' \Leftrightarrow$ There exists a $D \in \mathcal{E}$ such that $E \perp D$ and $E+D=F$

We say "$E$ and $F$ are *orthogonal*" for $E \perp F$. (OS2,4,5) imply that $0'=\mathbb{I}$ and $E''=E$ for $E \in \mathcal{E}$. A further relation $\prec$ is defined on $\mathcal{E}$ via: $E \prec F :\Leftrightarrow E \perp F'$ ($E,F \in \mathcal{E}$). Then $E \prec F$ if and only if $\mathcal{E}$ contains an element $D$ such that $D \perp E$ and $F=E+D$. Moreover, we have $0 \prec E \prec \mathbb{I}$ for all $E \in \mathcal{E}$.

The relation $\prec$ is reflexive by (OS4) or (OS5), but is not a partial ordering since it is neither anti-symmetric nor transitive in general. Therefore, the orthospace structure is far away from what is usually considered a quantum logic and is a rather weak structure the only purpose of which is to provide the opportunity to define states as an analogue of the classical probability measures. Further postulates concerning the states and conditional probabilities will be considered below and will then provide a sufficiently rich structure.

A *state* on an orthospace $\mathcal{E}$ is a map $\mu: \mathcal{E} \rightarrow [0,1]$ such that $\mu(\mathbb{I})=1$ and $\mu(E+F)=\mu(E)+\mu(F)$ for all orthogonal pairs $E,F \in \mathcal{E}$. Then $\mu(0)=0$, and $\mu$ is additive for each finite family of pairwise orthogonal elements in $\mathcal{E}$. (OS6) ensures that $\mu(E) \leq \mu(F)$ for $E \prec F$.

**Definition 2.1:** (i) *A* **σ-orthospace** *is an orthospace $\mathcal{E}$ such that $\Sigma_{n=1}^{\infty} E_n$ is defined in $\mathcal{E}$ for every sequence of mutually orthogonal events $E_n$.*
(ii) *A state $\mu$ on a σ-orthospace $\mathcal{E}$ is called* **σ-additive** *if $\mu(\Sigma_{n=1}^{\infty} E_n)=\Sigma_{n=1}^{\infty} \mu(E_n)$ for every sequence of mutually orthogonal events $E_n$.*

The elements $E \in \mathcal{E}$ are interpreted as *events* and will be called so in the following. Orthogonality means that the events exclude each other. The (only partially defined) operation + is interpreted as the *or* connection of mutually exclusive events, $E'$ is the *negation* of $E$. For a state $\mu$, the interpretation of the real number $\mu(E)$ is that of the *probability* of the event $E$ in the state $\mu$.

If $\mu$ is a state on an orthospace $\mathcal{E}$ and $E \in \mathcal{E}$ with $\mu(E)>0$ and if $\nu$ is another state such that $\nu(F)=\mu(F)/\mu(E)$ holds for all $F \in \mathcal{E}$ with $F \prec E$, then $\nu$ is called a *conditional probability* of $\mu$ under $E$. Essential shortcomings of this conditional probability are that such a state ν may not exist at all and that, if such a state exists, it may not be unique. The requirement that unique conditional probabilities must exist guides us to the following definition of σ-*UCP spaces* (which is the adaptation of the UCP spaces considered in [10] to σ-additive states).

**Definition 2.2:** *A* **σ-UCP space** *is a* σ-orthospace $\mathcal{E}$ *satisfying the following two axioms*:
(UC1) *If $E,F \in \mathcal{E}$ and $E \neq F$, then there is a* σ-additive state μ with $\mu(E) \neq \mu(F)$.
(UC2) *For each* σ-*additive state* μ *and* $E \in \mathcal{E}$ *with* $\mu(E) > 0$, *there exists one and only one* σ-*additive conditional probability* $\mu_E$ *of* μ *under* $E$.

$\mu_E(F)$ is the probability of the event $F$ in the state μ after the event $E$ has been observed. Using the same terminology as in mathematical probability theory, we will also write $\mu(F|E)$ for $\mu_E(F)$ in the sequel. If $\mu(E)=1$, then $\mu_E=\mu$ and $\mu(F|E)=\mu(F)$ for all $F \in \mathcal{E}$.

There is a σ-additive state μ with $\mu(E)=1$ for each element $E \neq 0$ in a σ-UCP space, since from (UC1) we get a σ-state ν with $\nu(E) > 0$, and then choose $\mu = \nu_E$.

(UC1) implies the uniqueness of $D$ in (OS6): $E+D_1=F=E+D_2$, then $\mu(E)+\mu(D_1)=\mu(F)=\mu(E)+\mu(D_2)$ for all σ-additive states μ, hence $\mu(D_1)=\mu(D_2)$ for all σ-additive states μ and $D_1=D_2$.

Moreover, if $E \prec F$ and $F \prec E$ for $E,F \in \mathcal{E}$, then $E=F$; i.e. the relation $\prec$ is anti-symmetric: If $F=E+D_1$ and $E=F+D_2$, then $\mu(F)=\mu(E)+\mu(D_1)=\mu(F)+\mu(D_2)+\mu(D_1)$, therefore $\mu(D_1)=\mu(D_2)=0$ for all σ-additive states μ, and $D_1=D_2=0$ by (UC1). Note that the relation $\prec$ need not be transitive so far. Furthermore: $E \perp E \Leftrightarrow E \perp \mathbb{I} \Leftrightarrow E=0$ (If $E \perp E$, then $E \perp E+E'=\mathbb{I}$ by (OS3,5). If $E \perp \mathbb{I}$, then $E \perp 0'$ and $E' \perp 0$, i.e. $E \prec 0$ and $0 \prec E$, hence $E=0$.).

In [10], the concepts of *statistical predictability* (state-independence of the conditional probability) and *compatibility* have been introduced. The adaptation of these concepts to σ-UCP spaces is straight forward, but not needed for the purpose of the present paper.

## 3. Observables

An observable is supposed to be the analogue of classical random variable which is a measurable point function $f$ between two measurable spaces. With the σ-UCP space model, there are no points, but only events. A closer look at classical probability theory shows that the map $X$ allocating the event $f^{-1}(E)$ to the event $E$ is more essential to the theory than $f$ itself. The map $X$ is a homomorphism between the σ-algebras of events. An observable is therefore defined as a homomorphism between two σ-orthospaces; similar but less general definitions of observables can be found in [4,12,16].

**Definition 3.1:** *A map $X$ from a* σ-orthospace $\mathcal{F}$ *to a* σ-orthospace $\mathcal{E}$ *is called an* **observable** *if*
(i) $X(\mathbb{I})=\mathbb{I}$,
(ii) $X(E) \perp X(F)$ *in* $\mathcal{E}$ *for all pairs $E,F \in \mathcal{F}$ with $E \perp F$, and*
(iii) $X\left(\sum_{n=1}^{\infty} F_n\right) = \sum_{n=1}^{\infty} X\left(F_n\right)$ *for every sequence of mutually orthogonal events $F_n$ in $\mathcal{F}$.*

Then $X(0)=0$ and $X(F')=X(F)'$ for every $F \in \mathcal{E}$. If μ is a σ-additive state on $\mathcal{E}$, a σ-additive state $\mu^X$ is defined on $\mathcal{F}$ via $\mu^X(F):=\mu(X(F))$; $\mu^X$ is called the distribution of $X$ under μ.

The classical σ-algebras and particularly the system $\mathcal{B}$ of Borel-measurable sets in $\mathbb{R}$ are σ-orthospaces. If $\mathcal{F}=\mathcal{B}$, the observable $X:\mathcal{B} \to \mathcal{E}$ is called a *real-valued* or $\mathbb{R}$-*valued* observable on $\mathcal{E}$, although it is a map from $\mathcal{B}$ to $\mathcal{E}$. The reason is that we want to keep the notation in line with what is called an real-valued classical random variable. Such an observable $X$ is *bounded* if

$$\|X\| := \inf\left\{r \geq 0 \,\big|\, X\left([-r,r]\right) = \mathbb{I}\right\}$$

is finite. Now let $O_b(\mathcal{E}, \mathbb{R})$ denote the set of all bounded $\mathbb{R}$-valued observables on $\mathcal{E}$. The *expectation value* of a real-valued observable $X$ in a σ-additive state μ on $\mathcal{E}$ is defined as

$$Exp_\mu(X) := \int t \, d\mu^X,$$

if the measure integral exists. The integral always exits if $X$ is bounded.

With a real-valued observable $X$ and a measurable function $f: \mathbb{R} \rightarrow \mathbb{R}$, another real-valued observable $Y$ is defined via $Y(B) := X(f^{-1}(B))$ for $B \in \mathcal{B}$; then $Exp_\mu(Y) := \int f(t) \, d\mu^X$ for any σ-additive state μ on $\mathcal{E}$. This observable $Y$ is denoted by $f(X)$ in the sequel. If $|f| \leq r$ for some $r \geq 0$, then $\|f(X)\| \leq r$. Thus $X^k$ and $sX$ are defined for any non-negative integer $k$ and any real number $s$, and we have $\|X^k\| = \|X\|^k$ and $\|sX\| = |s| \, \|X\|$. An observable $\chi_E \in O_b(\mathcal{E}, \mathbb{R})$ is allocated to each $E \in \mathcal{E}$ via

$$\chi_E(B) := \begin{cases} E & \text{for } 1 \in B \text{ and } 0 \notin B \\ E' & \text{for } 1 \notin B \text{ and } 0 \in B \\ 0 & \text{for } 1 \notin B \text{ and } 0 \notin B \\ \mathbb{I} & \text{for } 1 \in B \text{ and } 0 \in B \end{cases}$$

for $B \in \mathcal{B}$. Then $Exp_\mu(\chi_E) = \mu(E)$ for every σ-additive state μ on $\mathcal{E}$. Moreover, if $X$ is a real-valued observable, then $\chi_{X(B)} = I_B(X)$, where $B$ is any Borel set and $I_B$ is the indicator function with $I_B(t) = 1$ for $t \in B$ and $I_B(t) = 0$ for $t \notin B$.

The spectral measure of a self-adjoint operator on a complex or real Hilbert space provides a real-valued observable in the sense of the above definition. The one-to-one correspondence between the self-adjoint operators and their spectral measures is the reason why quantum-mechanical observables are usually understood as operators.

## 4. Three further axioms

A classical σ-algebra is a σ-UCP space with $\mu(E|F)\mu(F) = \mu(E \cap F) = \mu(F|E)\mu(E)$. In the quantum-mechanical Hilbert space model, we have $\mu(E|F)\mu(F) = Exp_\mu(FEF) \neq Exp_\mu(EFE) = \mu(F|E)\mu(E)$ (see [10]). In both cases, however, the conditional probabilities satisfy the equation $\mu(E|F)\mu(F) + \mu(E'|F')\mu(F') = \mu(F|E)\mu(E) + \mu(F'|E')\mu(E')$, which becomes our first axiom (A1) (see below).

Moreover, in both cases $\mu(E|F)\mu(F)$ is identical with the expectation value of a certain observable, this is the event $E \cap F$ in the first case and the (spectral measure of the) operator $FEF$ in the second case. This motivates the second axiom (A2).

In the last section, $X^k$ and $sX$ could be defined for a real-valued observable $X$, but the sum of two real-valued observables $X$ and $Y$ is not defined. An addition operation for observables is important for physical as well as mathematical reasons (e.g. for the formulation of a law of large numbers or a central limit theorem). This brings us to the third axiom (A3).

*Let $\mathcal{E}$ be a σ-UCP space.*

(A1) $\mu(E|F)\mu(F)+\mu(E'|F')\mu(F')=\mu(F|E)\mu(E)+\mu(F'|E')\mu(E')$ *for all events E and F and all σ-additive states μ on $\mathcal{E}$.*

(A2) *For each pair of events E and F there is a bounded real-valued observable $U_E(F)$ such that $\mu(F|E)\mu(E)=Exp_\mu(U_E(F))$ for every σ-additive state μ on $\mathcal{E}$.*

(A3) *For each pair of bounded real-valued observables X and Y there is one and only one bounded real-valued observable X+Y such that $Exp_\mu(X+Y)=Exp_\mu(X)+Exp_\mu(Y)$ for every σ-additive state μ on $\mathcal{E}$.*

Note that the sum of bounded $\mathbb{R}$-valued observables (spectral measures of bounded self-adjoint operators) exists in the quantum-mechanical model, but that the sum of bounded $\mathbb{C}$-valued observables (spectral measures of bounded normal operators) does not exist (since the sum of normal operators is not normal unless the operators commute). This means that the $\mathbb{R}$-valued observables play a distinguished role.

In the next three sections of the present paper, it will now be proved step-by-step that the system of bounded real-valued observables, equipped with this +-operation and a multiplication operation that will be defined later on, forms a Jordan algebra.

## 5. The addition operation

The axiom (A3) implies that the + operation on $O_b(\mathcal{E},\mathbb{R})$ is commutative as well as associative. Thus $O_b(\mathcal{E},\mathbb{R})$ becomes a real-linear space. The zero element is $0:=\chi_0$. Moreover, if $X,Y\in O_b(\mathcal{E},\mathbb{R})$ are such that $Exp_\mu(X)=Exp_\mu(Y)$ holds for all σ-additive states μ on $\mathcal{E}$, then $X=Y$ (which also implies the uniqueness of $U_E(F)$ when (A2) and (A3) both hold). If $g,f:\mathbb{R}\to\mathbb{R}$ are bounded measurable functions with $h:=g+f$, then $h(X)=g(X)+f(X)$ for every real-valued observable $X$.

**Lemma 5.1:** *Let $\mathcal{E}$ be a σ-UCP space.*

(i) $\|X\|=\sup\left\{\left|Exp_\mu(X)\right|:\mu \text{ is a σ - additive state on } \mathcal{E}\right\}$ *for $X\in O_b(\mathcal{E},\mathbb{R})$.*

(ii) *If* (A3) *holds, then $\|X+Y\|\leq\|X\|+\|Y\|$ and $\left\|X^2\right\|\leq\left\|X^2+Y^2\right\|$ for $X,Y\in O_b(\mathcal{E},\mathbb{R})$.*

*Proof*: (i) Obviously $|Exp_\mu(X)|\leq\|X\|$ for all σ-additive states μ on $\mathcal{E}$. Let ε>0. With $s:=\|X\|$ then either $X([s-\varepsilon,s])\neq0$ or $X([-s,-s+\varepsilon])\neq0$ and there is a σ-additive state μ on $\mathcal{E}$ such that either $\mu^X([s-\varepsilon,s])=1$ or $\mu^X([-s,-s+\varepsilon])=1$. In both cases we get $|Exp_\mu(X)|\geq\|X\|$ - ε.

(ii) follows immediately from (i). Use $Exp_\mu(Y^2)=\int s^2 d\mu^Y\geq0$ for the second inequality.

Now let $\mathcal{E}$ be a σ-UCP space that satisfies (A3). Then $\chi_{E+F}=\chi_E+\chi_F$ for any two orthogonal events $E$ and $F$ in $\mathcal{E}$. If $E_j$ $(1\leq j\leq k)$ are $k$ mutually orthogonal events in $\mathcal{E}$, the observable

$$\sum_{j=1}^{k} t_j\chi_{E_j}\in O_b(\mathcal{E},\mathbb{R})$$

with $t_j\in\mathbb{R}$ is called *primitive*. It is identical with the observable $X$ defined via

$$X(B) := \begin{cases} \sum_{t_i \in B} E_i & \text{for a Borel set } B \text{ with } 0 \notin B, \\ (\sum_{j=1}^{k} E_j)' + \sum_{t_i \in B} E_i & \text{for a Borel set } B \text{ with } 0 \in B. \end{cases}$$

Therefore

$$\left\| \sum_{j=1}^{k} t_j \chi_{E_j} \right\| = \max \left\{ |t_j| : 1 \leq j \leq k \right\}.$$

Note that the sum of two primitive observables is not primitive in general.

**Lemma 5.2:** *Let $\mathcal{E}$ be a σ-UCP space where* (A3) *holds.*
(i)   *The primitive observables are dense in the normed linear space $O_b(\mathcal{E}, I\!R)$.*
(ii)  *$\{X \in O_b(\mathcal{E}, I\!R) : \|X\| \leq 1\}$ is the closed convex hull of $\{ \chi_E - \chi_F : E, F \in \mathcal{E}\}$.*

*Proof*: (i) Let $X \in O_b(\mathcal{E}, I\!R)$ with $r := \|X\|$. Now approximate the function $f(s) := s$ on $[-r, r]$ uniformly by a sequence of step functions $f_n$ with a finite number of steps each; then $f_n(X)$ is a sequence of primitive observables with $\|X - f_n(X)\| \to 0$.

(ii) From **Lemma 5.1** (i) we get $\|\chi_E - \chi_F\| \leq 1$, and therefore $\|X\| \leq 1$ for every $X$ in the closed convex hull. Now assume $\|X\| \leq 1$ and approximate the function $f(s) := s$ on $[-1, 1]$ uniformly by a sequence of functions $f_n$ that are convex combinations of functions with values in $\{-1, 0, 1\}$. Then $f_n(X)$ is a sequence of observables in the convex hull of $\{ \chi_E - \chi_F : E, F \in \mathcal{E}\}$ and converges to $X$.

# 6. The multiplication operation

Postulating the existence of the product of two $I\!R$-valued observables $X$ and $Y$ in the same way as the sum is not possible since $Exp_\mu(XY) = Exp_\mu(X) Exp_\mu(Y)$ does not even hold in the classical case (unless $X$ and $Y$ are uncorrelated under μ). However, a multiplication operation will now be derived from (A1) and (A2).

**Theorem 6.1:** *Let $\mathcal{E}$ be a σ-UCP space that satisfies* (A1), (A2) *and* (A3). *Then there is a unique commutative multiplication operation $\circ$ with unit element $I\!I := \chi_{I\!I}$ on $O_b(\mathcal{E}, I\!R)$ such that*
(i)   *$X \circ (sY + tZ) = s(X \circ Y) + t(X \circ Z)$ for $X, Y, Z \in O_b(\mathcal{E}, I\!R)$ and $s, t \in I\!R$,*
(ii)  *$\|X \circ Y\| \leq \|X\| \|Y\|$ for $X, Y \in O_b(\mathcal{E}, I\!R)$,*
(iii) *$\chi_E \circ \chi_E = \chi_E$ for $E \in \mathcal{E}$, and $\chi_E \circ \chi_F = 0$ for $E, F \in \mathcal{E}$ with $E \perp F$.*
*Note that the multiplication operation $\circ$ is not associative in general.*

*Proof*: (1) Let $E \in \mathcal{E}$. From (A2) we get: $U_E(E) = \chi_E = U_E(I\!I) = U_{I\!I}(E)$, and $U_E(F) = 0$ for $F \in \mathcal{E}$ with $E \perp F$. We first define a certain extension of $U_E$ to linear combinations of the $\chi_F$ $(F \in \mathcal{E})$:

$$\widetilde{U}_E(Y) := \sum_{l=1}^{m} s_l U_E(F_l) \in O_b(\mathcal{E}, I\!R) \text{ for } Y = \sum_{l=1}^{m} s_l \chi_{F_l}.$$

Then

$$Exp_\mu \widetilde{U}_E(Y) = \mu(E) \sum_{l=1}^{m} s_l \mu_E(F_l) = \mu(E) Exp_{\mu_E} Y$$

for any σ-additive state μ on $\mathcal{E}$. Therefore, $\widetilde{U}_E$ is well-defined (i.e. independent of the special choice of the linear combination representing $Y$) and linear on the linear hull of $\{ \chi_F | F \in \mathcal{E}\}$.

Moreover, with **Lemma 5.1** (i), we get: $\|\tilde{U}_E(Y)\| \le \|Y\|$. We now define $\chi_E \circ Y$ for $E \in \mathcal{C}$ and $Y$ in the linear hull of $\{\chi_F | F \in \mathcal{C}\}$ via:

$$\chi_E \circ Y := \tfrac{1}{2}\big(Y + \tilde{U}_E(Y) - \tilde{U}_{E'}(Y)\big).$$

This immediately implies (iii), and moreover: $\chi_E \circ 0 = 0$ as well as $\chi_E \circ \mathbb{I} = \chi_E$. Furthermore, $\chi_E \circ Y$ is linear and continuous in $Y$ (with $E$ fixed).

(2) From (A1), we get for $E, F \in \mathcal{C}$:

$$U_E(F) + U_{E'}(F') = U_F(E) + U_{F'}(E')$$

and then

$$
\begin{aligned}
2\chi_E \circ \chi_F &= \chi_F + U_E(F) - U_{E'}(F)\\
&= \chi_F + U_F(E) + U_{F'}(E') - U_{E'}(F') - U_{E'}(F)\\
&= \chi_F + U_F(E) + \chi_{F'} - U_{F'}(E) - \chi_{E'}\\
&= \chi_E + U_F(E) - U_{F'}(E)\\
&= 2\chi_F \circ \chi_E .
\end{aligned}
$$

(3) We now define $X \circ Y$ for both $X$ and $Y$ in the linear hull of $\{\chi_F | F \in \mathcal{C}\}$ with $X = \sum_{j=1}^{k} t_j \chi_{E_j}$ as

$$X \circ Y := \sum_{j=1}^{k} t_j \big(\chi_{E_j} \circ Y\big).$$

We have to make sure that $X \circ Y$ does not depend on the special choice of the linear combination representing $X$. However, in the case $Y = \chi_F$ with $F \in \mathcal{C}$, we get from (2): $X \circ \chi_F = \chi_F \circ X$, which is well-defined by (1). Therefore $X \circ Y$ is well-defined and $X \circ Y = Y \circ X$ for all $Y$ that are linear combinations of such $\chi_F$. Moreover, $X \circ Y$ is continuous and linear in $Y$ with $X$ fixed as well as in $X$ with $Y$ fixed. We then get for $E_1, E_2 \in \mathcal{C}$:

$$\big(\chi_{E_1} - \chi_{E_2}\big) \circ Y = \tfrac{1}{2}\big(\tilde{U}_{E_1}(Y) - \tilde{U}_{E_1}(Y) + \tilde{U}_{E_2}(Y) - \tilde{U}_{E_2}(Y)\big).$$

Hence

$$\left\|\big(\chi_{E_1} - \chi_{E_2}\big) \circ Y\right\| \le 2\|Y\|,$$

and, using **Lemma 5.2** (ii):

$$\|X \circ Y\| \le 2\|X\|\|Y\|.$$

By LEMMA 5.2, the linear hull of $\{\chi_E | E \in \mathcal{C}\}$ is dense in $O_b(\mathcal{C}, \mathbb{R})$, and the multiplication operation $\circ$ has a unique continuous extension to $O_b(\mathcal{C}, \mathbb{R})$ such that (i) and (iii) are satisfied, but $X \circ Y$ may lie in the completion of $O_b(\mathcal{C}, \mathbb{R})$ and not in $O_b(\mathcal{C}, \mathbb{R})$ itself for $X, Y \in O_b(\mathcal{C}, \mathbb{R})$.

(4) For a primitive observable $X$, the product $X \circ X$ is identical with $X^2$ defined earlier as $f(X)$ with $f(s) := s^2$. This follows from (i) and (iii). Therefore, $X \circ X = X^2$ for all $X \in O_b(\mathcal{C}, \mathbb{R})$; use the same approximation of $X$ by primitive observables as in the proof of **Lemma 5.2** (i) and the continuity of the multiplication operation $\circ$. Then

$$X \circ Y = \tfrac{1}{2}(X + Y)^2 - \tfrac{1}{2}X^2 - \tfrac{1}{2}Y^2 \in O_b(\mathcal{C}, \mathbb{R})$$

for $X, Y \in O_b(\mathcal{C}, I\!\!R)$. Furthermore, the multiplication operation $\circ$ is uniquely determined by this equation.

(5) Since we have $Exp_\mu(X^2)=\int s^2 d\mu^X \geq 0$, the Cauchy-Schwarz inequality holds for the bilinear form $X, Y \rightarrow Exp_\mu(X \circ Y)$ on $O_b(\mathcal{C}, I\!\!R)$, where $\mu$ is any $\sigma$-additive state on $\mathcal{C}$. Then for $X, Y \in O_b(\mathcal{C}, I\!\!R)$:

$$\left( Exp_\mu \left( X \circ Y \right) \right)^2 \leq Exp_\mu \left( X^2 \right) Exp_\mu \left( Y^2 \right),$$

and by **Lemma 5.1** (i):

$$\| X \circ Y \|^2 \leq \| X^2 \| \| Y^2 \| = \| X \|^2 \| Y \|^2,$$

from which we get (ii).

With the multiplication operation of **Theorem 6.1**, we have $f(X) \circ g(X)=h(X)$ with $h:=fg$ for any real-valued observable $X$ and any bounded measurable functions $f,g$ on $I\!\!R$. This can easily be proved by applying the equation $2X \circ Y=(X+Y)^2-X^2-Y^2$.

One may decline to take this equation as a definition of the product $\circ$, using the definition of the square from section 3 and to drop (A1) and (A2) as well as (UC2). This is the so-called Segal product[13,14]. Then, however, the distributive law ((i) in 6.1) cannot be proved[15], which is the reason why other authors postulate the distributive law for the product as an extra axiom although a physically or statistically plausible justification for this axiom is hard to fine.

# 7. The Jordan property

A Jordan algebra satisfies the condition $X \circ (X^2 \circ Y)=X^2 \circ (X \circ Y)$ for all elements $X$ and $Y$ in the algebra. If a real algebra has a finite dimension and satisfies some conditions which hold in $O_b(\mathcal{C}, I\!\!R)$, the Jordan condition becomes equivalent to the condition that each element of the algebra lies in an associative sub-algebra[8]. In $O_b(\mathcal{C}, I\!\!R)$, an associative sub-algebra containing a given $X$ is $\{f(X): f$ is a bounded measurable function on $I\!\!R$ $\}$. Therefore, $O_b(\mathcal{C}, I\!\!R)$ is a Jordan algebra if its dimension is finite. We shall now show that $O_b(\mathcal{C}, I\!\!R)$ is a Jordan algebra in the infinite-dimensional case as well, using the methods of [8] where applicable. In [8], the finite dimension is mainly needed to derive a spectral theorem. We are in the lucky situation to have such a theorem already; this is **Lemma 5.2** (i). Since observables are a kind of abstract spectral measures, spectral theory becomes quite simple in our case.

**Lemma 7.1:** *Under the assumptions of* **Theorem 6.1***, the identity* $\chi_E \circ (\chi_F \circ Y)= \chi_F \circ (\chi_E \circ Y)$ *holds for any two orthogonal events* $E,F \in \mathcal{C}$ *and any* $Y \in O_b(\mathcal{C}, I\!\!R)$*; i.e.* $\chi_E$ *and* $\chi_F$ *operator-commute*[7].

For the proof of this lemma it is referred to [8]. $O_b(\mathcal{C}, I\!\!R)$ satisfies all the assumptions needed there, and the finite dimension is not relevant for this proof.

**Theorem 7.2:** *Under the assumptions of* **Theorem 6.1***,* $O_b(\mathcal{C}, I\!\!R)$ *is a Jordan algebra.*

*Proof*: Due to **Lemma 5.2** (i), it is sufficient to prove the identity $X \circ (X^2 \circ Y)=X^2 \circ (X \circ Y)$ for $X, Y \in O_b(\mathcal{C}, I\!\!R)$ with $X$ being primitive. We therefore consider

$$X = \sum\nolimits_{k=1}^{n} t_k \chi_{E_k}$$

with mutually orthogonal events $E_k$ and $t_k \in I\!\!R$ $(1 \leq k \leq n)$. Then by **Lemma 7.1**:

$$X^2 \circ (X \circ Y) = \sum_{k=1}^{n}\sum_{l=1}^{n} t_k{}^2 t_l \chi_{E_k} \circ (\chi_{E_l} \circ Y) = \sum_{k=1}^{n}\sum_{l=1}^{n} t_k{}^2 t_l \chi_{E_l} \circ (\chi_{E_k} \circ Y) = X \circ (X^2 \circ Y).$$

The map $E \rightarrow \chi_E$ provides an isomorphism form $\mathcal{E}$ onto the system of idempotent elements in $O_b(\mathcal{E}, \mathbb{R})$, the completion of which becomes a so-called JB algebra[7]. This finally implies that $\prec$ is an orthomodular partial ordering. Moreover, since almost all JB algebras can be represented as a Jordan algebra of self-adjoint operators on a Hilbert space[1,7], we thus arrive very closely at the standard Hilbert space model of quantum mechanics.

If the multiplication operation on $O_b(\mathcal{E}, \mathbb{R})$ is associative, then $O_b(\mathcal{E}, \mathbb{R})$ is isomorphic to an algebra of real-valued functions[7] and $\mathcal{E}$ is a ($\sigma$-complete) Boolean lattice, i.e. an associative multiplication operation reduces to the classical case.

## 8. Conclusions and remarks

We have seen that the $\sigma$-UCP spaces and the axioms (A1), (A2) and (A3) presented above provide an axiomatic approach to quantum mechanics, incorporating a statistical interpretation from the very beginning and leading to real Jordan algebras. The structure theory of Jordan operator algebras finally provides the link to the conventional Hilbert space or C*-/W*-formalism of quantum mechanics.

This approach includes the physically relevant type II,III von-Neumann algebras which are not covered by the purely logical approaches and it excludes some physically irrelevant cases that the purely logical approaches are unable to rule out. It is closer to Kolmogorov's measure-theorectic access to classical probability theory than other approaches. The existence of the distributive multiplication operation for the bounded real observables need not be postulated without a satisfying justification, but is derived from other postulates concerning certain properties of the conditional probabilities and observables.

The connection between conditional probabilities (though the definition does not coincide with our one) and real Jordan algebras was discovered by Gunson[5]. His results were improved by Guz[6], but only the finite events (sum of a finite number of orthogonal minimal events) and their orthogonal complements could be embedded in a Jordan algebra.

Guz also proposed an algebraic approach where two of his eleven axioms coincide with (A1) and (A2). Axiom (A1) appeared for the first time and its possible interpretation is discussed in Alfsen and Shultz's paper[2]. Better known is axiom (A3) which is the major ingredient for the definition of the so-called *sum logics*[12].